\documentstyle[twoside,fleqn,espcrc2,epsf]{article}

\newcommand{\AmS}{{\protect\the\textfont2
  A\kern-.1667em\lower.5ex\hbox{M}\kern-.125emS}}

\title{
\vspace{-8mm}
\rightline{\small HUB--EP--97/66}
\vspace{-2mm}
\rightline{\small September 19, 1997}
Topology without cooling: 
instantons and monopoles \\
   near to deconfinement
         \thanks{combining a talk given by E.-M. Ilgenfritz
	 and a poster presented by S. Thurner} 
}

\author{M. Feurstein$^{\scriptsize a}$, 
E.-M. Ilgenfritz$^{\scriptsize b}$\thanks{supported by 
                                          DFG under grant Mu932/1-4}, 
H. Markum$^{\scriptsize a}$,
M. M\"uller-Preussker$^{\scriptsize b}$
and S. Thurner$^{\scriptsize a}$ \\
\vspace{3mm}
$^{\scriptsize a}$ Institut f\"{u}r Kernphysik, TU Wien,
         Wiedner Hauptstra\ss e 8-10, A-1040 Vienna, Austria\\ 
\vspace{3mm}
$^{\scriptsize b}$ Institut f\"{u}r Physik, Humboldt-Universit\"at zu Berlin,
         Invalidenstra\ss e 110, Berlin, Germany\\
}

\begin{document}

\begin{abstract}
In an attempt to describe the change of topological 
structure of pure $SU(2)$ gauge theory near deconfinement 
a renormalization group inspired method is tested.
Instead of cooling, blocking and subsequent inverse blocking 
is applied to Monte Carlo
configurations to capture topological features at a 
well-defined scale. 
We check that this procedure largely conserves 
long range physics like string tension.
UV fluctuations and lattice artefacts are removed which otherwise 
spoil topological charge density and Abelian monopole currents.
We report the behaviour of topological susceptibility and monopole
current densities across the deconfinement transition and relate 
the two faces of topology to each other.
First results of a cluster analysis are described.
\end{abstract}

\maketitle
\section{INTRODUCTION}
An important ingredient of the construction of perfect actions 
\cite{p.hasenfratz}
is inverse blocking. 
This method gives smooth 
interpolations of background lattice gauge fields on finer and finer
lattices which 
are free of quantum fluctuations. 
Exploring the topological structure of lattice gauge  
vacua (and thermal states) requires 
such an interpolation.
It is well-known that some 
smoothing is needed just to determine the global  
topological charge in an unambiguous way \cite{cool85}.
Cooling has been proposed iteratively transforming
Monte Carlo (MC) configurations into smoother ones by  
{\it unconstrained
relaxation} w.r.t. to some action. Due to the 
diffusive character of cooling it is hard to predict after how many 
iterations
cooled fields exhibit
structures 
characteristic for the quantum ensemble 
at a definite scale.  
A random walk estimate asserts that $n$ cooling iterations will keep  
structures intact at distance 
$\frac{r}{a} \simeq \sqrt{n}$. Relying on this, cooling
is used {\it e. g.} to measure
the connected field strength correlator at intermediate distances 
\cite{digiacomo}.
There are attempts to improve cooling by use of 
improved actions \cite{stamatescu} (measuring with improved topological density)
including extended Wilson loops. This can protect instantons
at a scale of a few lattice spacing, but still
instanton-antiinstanton pairs annihilate
at some stage of improved cooling.  

Inverse blocking captures
topological structure near to the scale of the original lattice spacing 
by a strictly local
procedure. Thus, perfect actions prove their 
{\it diagnostic power} based on the renormalization group. 
DeGrand et al.\cite{degrand} 
demonstrated that one step of inverse blocking
suffices to define a perfect total 
topological charge even if one deals with instantons with $\rho$
not bigger than 
the lattice spacing.
This result has inspired us to examine further the 
capability of inverse blocking.
Our aim was to detect more details of topological 
structure in terms of {\it topological charge density and monopole
currents} 
and to study how they change at the deconfining transition \cite{wir}. 

\section{FIXED POINT ACTION AND \\
	     INVERSE BLOCKING}
We have used a truncated fixed point
action with plaquettes and tilted $3$-dimensional 
$6$-link loops in various representations \cite{degrand,wir} 
\begin{equation}
S_{FP}(U)=\sum_{type~i} \sum_{C_{i}} \sum_{j=1}^{4}
w(i,j) (1 - {1 \over 2}~{\mathrm tr}~U_{C_{i}} )^j .
\end{equation}
The lattice size was $L_s=12$ and $L_t=4$ where simulations 
have been carried out at
$\beta=1.40$, $1.50$, $1.54$, $1.60$ and
$1.80$. This interval encloses 
$\beta_c = 1.575(10)$ \cite{degrand} where 
the deconfinement transition happens for
$L_t=4$.
Our finite temperature results presented below are based on  
$300$ MC (unsmoothed) configurations 
and $100$ smoothed ones  
per $\beta$ value.

Given a coarse configuration with links $V$,
inverse blocking gives the {\it constrained minimum} of 
the fixed point action as a configuration $U$ on the next finer
lattice. 
The constraint is formulated 
including a blocking kernel $T(U,V)$ into 
the full action functional to be minimized
\begin{equation}\label{eq:full_action}
S_{full}(U,V) ~= ~S_{FP}(U) ~+ ~\kappa ~T(U,V) .
\end{equation}
Classical perfectness requires saturation
\begin{equation}\label{eq:saturation}
\mbox{Min}_U(S_{full}(U,V)) ~ = ~ S_{FP}(V) .
\end{equation}
We found this saturation to be fulfilled for 
equilibrium configurations 
with an accuracy of a few percent
(depending on $\beta$)
using an effective
$\kappa=5.15$. We convinced ourselves that the physical
results to be presented do not very much depend on this
parameter.

With a perfect action it should not
matter at which level the MC code
is actually running. 
We have simulated on the fine lattice creating links $U$, 
blocked them to coarse lattice links $V$.
Keeping $V$ fixed, by inverse blocking (IB) we obtained $U_{smooth}$, 
the smoothest interpolation to $V$. 
Although the final analysis is done on the 
fine lattice, structures that are resolved belong to the coarse
lattice.
Being of non--perturbative, long range origin, a confining potential  
should survive the smoothing step as far as no structures
below the coarse level contribute to the string tension.
Contrary to this, big changes
of topological and monopole densities 
are expected
since quantum fluctuations 
and dislocations
(otherwise counted on the fine lattice)
are washed out by smoothing. 

\section{DOES THE STRING TENSION \\
	 SURVIVE SMOOTHING ?}
For calibration of the finite 
temperature lattices for the set of $\beta$
values
we have measured the zero temperature string tension on a symmetric 
lattice $12^4$. We have considered 
temporal, spatially fuzzy Wilson loops 
and
extracted the string tension from fits to the potential energy. 
We have found a window of approximate scaling around $\beta_c$.
The two--loop expression for $a(\beta)\Lambda_{L}$
has been used, giving $\sigma/\Lambda_{L}^2$ 
from data for $\sigma a^2$ (see table \ref{tab:loop_string_tens}).
\begin{table}[h]
\begin{center}
\begin{tabular}{|l|c|c|c|}
\hline
$\beta$   & $1.50$     & $1.54$     & $1.60$      \\ 
\hline
\hline
MC & $135(2)$    & $129(2)$    & $111(2)$    \\
IB & $100.5(10)$ & $104.5(20)$ & $96.8(20)$  \\
\hline
\end{tabular}
\end{center}
\caption{\sl String tension 
	    $\sigma/\Lambda_{L}^2$ 
	    at $T=0$ 
	    from fuzzy Wilson loops 
	    on a $12^4$ lattice}
\label{tab:loop_string_tens}
\end{table}
\vspace*{-1.5cm}
\begin{figure}[!thb]
\centering
\epsfxsize=7.5cm\epsffile{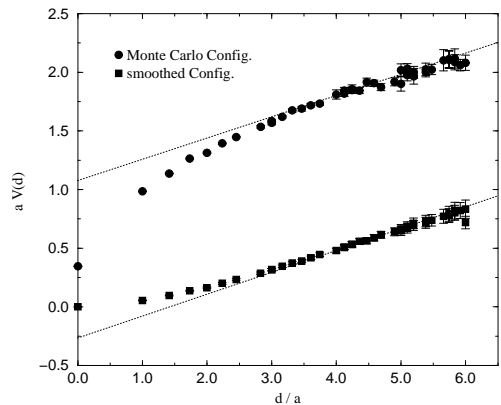}
\vspace*{-0.8cm}
\caption{$\overline{q} q $ \sl potential from Polyakov line correlators 
for MC (above) and smoothed (IB) configurations at 
$\beta=1.5$}
\label{fig:potential}
\end{figure}
From $75$ to $90$ per cent (with increasing $\beta$)
of the $T=0$ string tension  
are preserved after smoothing. 
The loss has to be attributed to
confining excitations smaller than the spacing $2a$ of the coarse lattice.

A similar effect on the string tension $\sigma(T)$ is observed
(see table \ref{tab:pol_string_tens}).
\begin{table}[h]
\begin{center}
\begin{tabular}{|l|c|c|}
\hline
$\beta$     &   $1.50$    & $1.54$ \\ 
\hline
$T/T_{c}$   &   $0.834$   & $0.919$ \\
\hline
\hline
MC &   $110(4)$   & $118(4)$ \\
IB &   $104(2)$   & $92(2)$ \\
\hline
\end{tabular}
\end{center}
\caption{\sl String tension 
	    $\sigma(T)/\Lambda_L^2$ 
			    from Polyakov line 
correlators at two temperatures below $T_c$ 
}
\label{tab:pol_string_tens}
\end{table}
In Fig. \ref{fig:potential} 
the logarithm of the
Polyakov line correlator 
$\langle L({\bf 0}) L({\bf x}) \rangle 
\propto \exp (- V_{\overline{q} q}({\bf x})/T)$
is shown  
for $\beta=1.5$ 
where $V_{\overline{q} q}({\bf x}) \sim \sigma(T) |{\bf x}|$ at 
large distances. With short range quantum fluctuations present in the
MC configurations, $V_{\overline{q} q}({\bf x})$ 
contains also
the
Coulomb potential.  
As expected, the latter is removed by smoothing while the
string tension $\sigma(T)$  
is moderately reduced
(by less than $5$ per cent at $T < 0.8~T_{c}$).
Very near to the deconfinement temperature, the string
tension becomes weaker if measured on
smoothed configurations.

\section{TOPOLOGICAL VS. MONOPOLE \\
	  DENSITY} 
Monopole currents $m_{\mu}(x)$ are detected
in the Abelian projection of $SU(2)$ gauge
configurations after these have been put into the maximally Abelian gauge.
We reconstruct the topological charge density
calculating the
local contributions to L\"uscher's geometrical charge 
and the naive (plaquette oriented) definition of Pontryagin density.  
Thus we could cross-check
both definitions.
For MC configurations, the topological charge density 
is obscured by short range quantum fluctuations and dislocations in both cases. 
Concerning number and locations of monopole currents the
situation is similar.
Evaluating the correlator 
between $|q(x)|$ (of the
gauge invariant topological density) and the 
(gauge dependent !) 
monopole density $|m_{\mu}(x)|$ 
fluctuations are averaged out such that {\it no
smoothing} is required.
This was observed earlier for   
MC configurations generated with Wilson action \cite{vienna}.
\begin{figure}[!thb]
\centering
\epsfxsize=7.5cm\epsffile{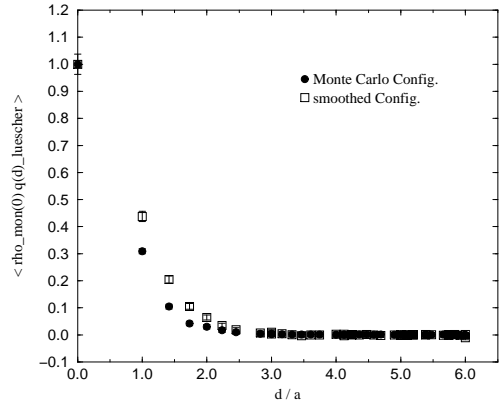}
\vspace*{-0.8cm}
\caption{\sl Normalized correlation between monopole
and topological density
at $\beta=1.5$ for MC and smoothed configurations}
\label{fig:qlumon_corr}
\end{figure}
We show in Fig. \ref{fig:qlumon_corr} 
the (normalized) correlator comparing MC configurations 
generated with the fixed point action at $\beta=1.5$ 
(confinement phase) with the
corresponding ensemble of smoothed
configurations where the correlator is slightly wider.
The same correlation functions 
at $\beta=1.8$ in the deconfined phase (not shown here),
are found to be the same
within error bars.  
The pointlike Abelian monopoles seem to be
accompanied by a cloud of topological density, irrespective of
the phase. The size of this cloud
changes proportional to $1/T$.
This correlation is present  
in MC configurations 
and in their smoothed counterparts.

\begin{figure}[!thb]
\centering
\epsfxsize=7.5cm\epsffile{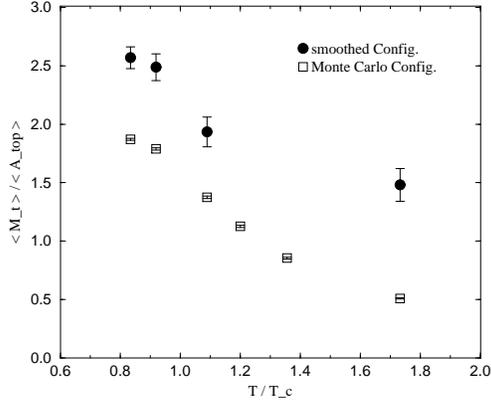}
\vspace*{-0.8cm}
\caption{\sl Ratio between timelike monopole number 
$\langle M_{t} \rangle$
and topological activity $\langle A_{top} \rangle$ vs. $T/T_{c}$ for MC
and smoothed configurations}
\label{fig:MTdurchA}
\end{figure}
\begin{figure}[!thb]
\centering
\epsfxsize=7.5cm\epsffile{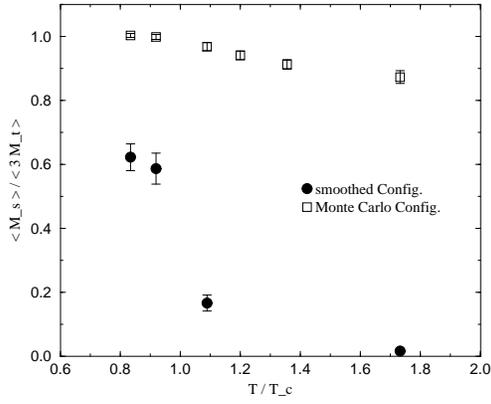}
\vspace*{-0.8cm}
\caption{\sl Asymmetry between timelike and spacelike 
monopole currents vs. $T/T_c$
for MC and smoothed configurations}
\label{fig:monasymmetrie}
\end{figure}
This observation suggests that a global relation 
exists between the topological activity
$A_{top}=\sum_x |q(x)|$
and the number of monopole currents
$M_{\mu}=\sum_x|m_{\mu}(x)|$, almost independent of smoothing {\it and} phase. 
Although both quantities are strongly reduced
by smoothing, the ratio of averages 
does not change very much and remains a smooth
function of temperature. This becomes even clearer if one considers 
only the temporal monopole current $M_{t}$, as 
Fig. \ref{fig:MTdurchA} demonstrates.  
In view of the different role timelike ($M_{t}=M_4$)
and spacelike monopole currents
play for confinement (for {\it magnetic} confinement {\it above} $T_c$),
the ratio between $M_s=\sum_1^3M_{\mu}$   
and $3~M_t$ is expected to be an disorder parameter
of the deconfinement transition \cite{mmp:asymm}. 
The high monopole activity in MC
configurations hides this while smoothing makes it visible, 
as the comparison 
in Fig. \ref{fig:monasymmetrie} demonstrates.

\section{TOPOLOGICAL SUSCEPTIBILITY}
The topological susceptibility is defined here by the fluctuation of 
topological charge per configuration, 
$\chi = \langle Q^2 \rangle /(N_{sites} ~a^4)$,
with charges $Q$
according
to the naive and L\"uscher's topological charge density. 
For MC configurations the two charges are almost uncorrelated
in contrast to smoothed ones.   
\begin{figure}[!thb]
\centering
\epsfxsize=7.5cm\epsffile{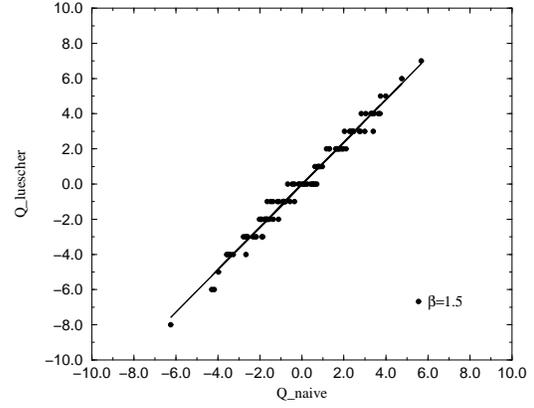}
\vspace*{-0.8cm}
\caption{\sl L\"uscher's vs. naive 
topological charge for smoothed configurations at $\beta=1.5$}  
\label{fig:Z_Faktor_scatter}
\end{figure}
\begin{table}[h]
\begin{center}
\begin{tabular}{|c|c|c|c|}
\hline
     $\beta$ & $T/T_{c}$ & $\chi^{(l)}/\Lambda_L^4$ & $\chi^{(n)}/\Lambda_L^4$ \\
\hline
\hline
     $1.50 $ & $0.834$ & $.287(24) 10^4$ & $.370(27) 10^2$ \\
     $1.54 $ & $0.919$ & $.310(27) 10^4$ & $.449(43) 10^2$ \\
     $1.61 $ & $1.089$ & $.235(19) 10^4$ & $.753(59) 10^2$ \\
     $1.80 $ & $1.732$ & $.190(16) 10^4$ & $.308(24) 10^3$ \\
\hline
     $1.50 $ & $0.834$ & $.358(47) 10^3$ & $.240(29) 10^3$ \\
     $1.54 $ & $0.919$ & $.324(43) 10^3$ & $.239(30) 10^3$ \\
     $1.61 $ & $1.089$ & $.138(25) 10^3$ & $.962(18) 10^2$ \\
     $1.80 $ & $1.732$ & $.164(11) 10^2$ & $.100(70) 10^2$ \\
\hline
\end{tabular}
\end{center}
\caption{\sl Topological susceptibilities of the MC (above) 
	     and smoothed (below) samples
}  
\label{tab:suscept}
\end{table}
Fig. \ref{fig:Z_Faktor_scatter} shows 
a corresponding scatter plot. From the slope we can extract a 
renormalization factor 
$Z_q(\beta)$ 
of the naive charge for smoothed configurations, 
$Z_q ~q(x) ~a^4 = q_{naive}(x) ~a^4$.
In the neighbourhood of the phase transition we obtained
$Z_q=0.826$, $.877$ and $.843$ for $\beta=1.50$, $1.54$ and $1.61$, 
respectively.

Susceptibility data for MC and smoothed configurations 
(still without renormalization of the naive topological charge) 
are given in table \ref{tab:suscept}. 
$\chi^{(l)}$ (based on L\"uscher's geometric charge) is reduced by smoothing 
by one to 
two orders of magnitude (going from confinement to deconfinement). 
This means that the fixed point action does not sufficiently suppress 
topological dislocations. They
are
smoothed away by blocking and inverse blocking.
This is in accordance to
the observation \cite{degrand} that a sensible
measurement of topological charge (even a geometrical one) 
requires  
interpolation by inverse blocking.

\begin{figure}[!thb]
\centering
\epsfxsize=7.5cm\epsffile{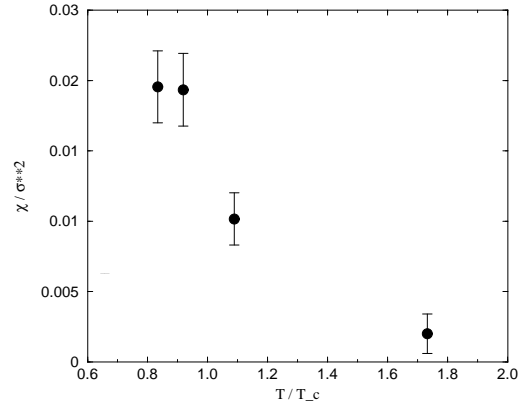}
\vspace*{-0.8cm}
\caption{\sl Topological susceptibility for smoothed configurations
vs. $T/T_c$}
\label{fig:suscept_ib}
\end{figure}

The naive topological density
has a {\it perturbative} $Z_q << 1$ 
for $SU(2)$ Wilson action in the relevant $\beta$ range.
Also with the fixed point action used for production of MC
configurations, the
naive topological susceptibility $\chi^{(n)}$ (without
proper renormalization) is 
two
orders of magnitude lower than $\chi^{(l)}$ defined
through (integer valued) 
charges if calculated for MC configurations.
After smoothing, however, 
the susceptibilities $\chi^{(n)}$ and $\chi^{(l)}$
do not differ by more than $30$ per cent.
This
difference can be accounted for by the effective renormalization factor 
defined
above for smoothed configurations.  
A unique $\chi$ emerges 
whose $\beta$-dependence is turned into $T$-dependence 
in Fig. \ref{fig:suscept_ib}.   
For definiteness, the lattice
spacing has been 
expressed through the zero temperature 
string tension measured
for each $\beta$, respectively,
on the symmetric lattice.
To normalize we can
assume a string tension $\sigma=(440 \mbox{MeV})^2$ and obtain 
a susceptibility $\chi=(165.5 \mbox{MeV})^4$
at $T=0.834~T_c$.
 
\section{ARE INSTANTONS BUILDING THE \\
	 TOPOLOGICAL SUSCEPTIBILITY ?}
A non-trivial two-point correlator of topological density
is difficult to obtain without smoothing. 
For MC configurations it vanishes at non--zero distance (apart from
a negative kinematical signal at $d<2a$). 
For smoothed configurations, however, the correlator can be described by
folding the instanton profile according to a dilute gas picture.
An ''instanton radius'' $\rho$ can be formally defined in this way. This
interpretation should be taken with care.
In confinement, {\it this} $\rho/a$ is found independent of $\beta$, 
near to the lowest
instanton size detectable by inverse blocking, somewhat
smaller than the lattice spacing of the blocked lattice. In deconfinement,
$\rho/a$ becomes
even smaller with increasing $\beta$. 

\begin{figure}[!thb]
\centering
\epsfxsize=7.5cm\epsffile{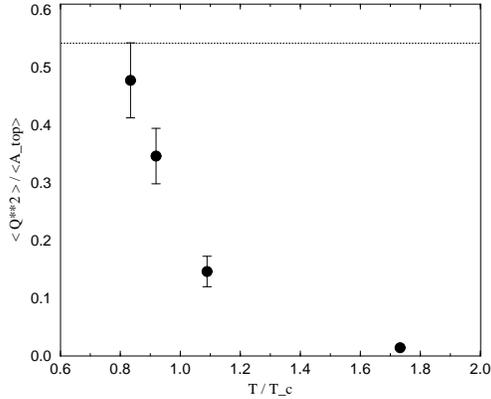}
\vspace*{-0.8cm}
\caption{\sl Ratio between $\langle Q^2 \rangle$ and
topological activity $\langle A_{top} \rangle$ 
of smoothed configurations 
vs. $T/T_{c}$}
\label{fig:QsqdurchA}
\end{figure}

Instanton models \cite{shuryak} use statistical mechanics arguments to relate
the density of
instantons and antiinstantons to vacuum energy and topological susceptibility
by low energy theorems 
(as well as to make predictions on the
multiplicity distribution).
Assumptions on instanton interactions enter into these models
which
can be checked measuring
the average density of instantons
{\it independently}
from the topological
susceptibility. A first attempt was to compare 
$A_{top}$ defined above (this makes sense only for smoothed configurations)
with $\chi$.
Over the temperature range considered $A_{top}$ decreases, but less 
rapid than $\chi$ (as can be seen in
Fig. \ref{fig:QsqdurchA}). The horizontal line is a bound set by
a low energy theorem for $T=0$
\begin{equation}\label{eq:1/N}
\chi = \frac{6}{11} \frac{\langle A_{top} \rangle}
				   {N_{sites} a^4} 
	         =  \frac{6}{11} \langle n_{+} + n_{-} \rangle  .
\end{equation}
\begin{figure}[!thb]
\centering
\epsfxsize=7.5cm\epsffile{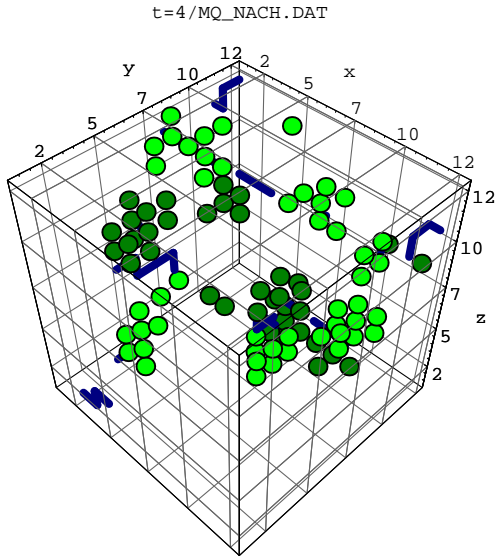}
\caption{\sl One timeslice of a smoothed configuration: 
monopole world lines and clusters of topological charge density 
with $|q^{(l)}(x)|>q_0$}
\label{fig:smooth_luescher}
\end{figure}
The crucial factor $4/b=12/(11N_c)$ 
($b$ is the one--loop coefficient in the $\beta$-function)
appears also in the instanton liquid equation of state 
$\langle N \rangle/V=\langle n_{+}+n_{-} \rangle=(4/b)|\varepsilon_{vac}|$,
the width of the multiplicity distribution 
$\langle N^2 \rangle - \langle N \rangle^2=
(4/b)\langle N \rangle$ (compressibility smaller than Poisson)
and in the entropic bound $S/(2\pi^2) > 4/b$ 
(for carriers of charge $Q=1$) banning dislocations.
Interactions leading to a $O(1/N_c)$ suppression of the topological
susceptibility compared to the densities $n_{\pm}$ 
have been discussed 
\cite{1/N}
in the earliest days of the instanton liquid model.  

In fact, $A_{top}$ is only an upper bound for $N$. 
Further illumination is expected from a cluster analysis of smoothed
configurations. A typical example is shown in Fig. \ref{fig:smooth_luescher}.
But there come surprises: 
Clusters do not look like classical instantons. The size cannot estimated
from the maximum of $|q(x)|$. If isolated charges are removed by few
cooling steps,
the 
cluster charges are centered around $Q_{cl}=\pm 1$. Although the
mean multiplicity of clusters drops drastically at $T_c$,
the variance follows Poisson's law at all temperatures across the transition.
We hope to clarify further the nature of the topological composition of
the $SU(2)$ vacuum in ongoing work.

\end{document}